\documentclass[aps,prl,twocolumn,superscriptaddress,showpacs]{revtex4-1}

\usepackage{graphicx}
\usepackage{amsmath}
\usepackage{longtable}
\usepackage{color}

\newcommand{\ncmox}{Na$_{1-x}$Ca$_x$Mn$_7$O$_{12}$}
\newcommand{\ncmo}{Na$_{0.75}$Ca$_{0.25}$Mn$_7$O$_{12}$}
\newcommand{\nmo}{NaMn$_7$O$_{12}$}
\newcommand{\cmo}{CaMn$_7$O$_{12}$}
\newcommand{\amthree}{$A$-Mn$^{3+}$}
\newcommand{\bmthree}{$B$-Mn$^{3+}$}
\newcommand{\bmfour}{$B$-Mn$^{4+}$}

\begin{document}

\title{Evolution of magneto-orbital order upon $B$-site electron doping in \ncmox\ quadruple perovskite manganites}

\author{R. D. Johnson}
\email{roger.johnson@physics.ox.ac.uk}
\affiliation{Clarendon Laboratory, Department of Physics, University of Oxford, Oxford, OX1 3PU, United Kingdom}
\author{F. Mezzadri}
\affiliation{Istituto dei Materiali per Elettronica e Magnetismo, CNR, Area delle Scienze, 43100 Parma, Italy}
\author{P. Manuel}
\author{D. D. Khalyavin}
\affiliation{ISIS Facility, Rutherford Appleton Laboratory-STFC, Chilton, Didcot OX11 0QX, United Kingdom}
\author{E. Gilioli}
\affiliation{Istituto dei Materiali per Elettronica e Magnetismo, CNR, Area delle Scienze, 43100 Parma, Italy}
\author{P. G. Radaelli}
\affiliation{Clarendon Laboratory, Department of Physics, University of Oxford, Oxford, OX1 3PU, United Kingdom}

\date{\today}

\begin{abstract}
We present the discovery and refinement by neutron powder diffraction of a new magnetic phase in the \ncmox\ quadruple perovskite phase diagram, which is the incommensurate analogue of the well-known \emph{pseudo}-CE phase of the simple perovskite manganites.  We demonstrate that incommensurate magnetic order arises in quadruple perovskites due to the exchange interactions between $A$ and $B$ sites. Furthermore, by constructing a simple mean field Heisenberg exchange model that generically describes both simple and quadruple perovskite systems, we show that this new magnetic phase unifies a picture of the interplay between charge, magnetic and orbital ordering across a wide range of compounds.

\end{abstract}

\maketitle

Since the early neutron scattering works by Wollan and Koehler \cite{Wollan1955} and their interpretation by Goodenough \cite{Goodenough1955}, manganese oxides with structures related to perovskite (the manganites) have represented prototypical systems to study the interplay between charge, orbital and magnetic order in the solid state.  Simple perovskites with general formula $A^{3+}_{1-x}A^{2+}_x$MnO$_3$ have been extensively studied, because such heterovalent $A$ site cation substitution enables parameters such as electron doping, bandwidth and disorder to be tuned over wide ranges \cite{tokura06}. The general phenomenology of these materials had been broadly understood by the end of the 1990's. For example, heavily distorted, narrow-bandwidth materials such as Pr$_{1-x}$Ca$_x$MnO$_3$ are always insulating in the absence of an applied magnetic field, and display ordering of charges, spins and orbitals with complex but \emph{commensurate} superstructures (Fig. 1a, adapted from \cite{tomioka96}). By contrast, for intermediate and wide bandwidth materials, a ferromagnetic metallic region is present at low temperatures and intermediate doping, competing with the insulating state.  This competition can be tuned by an external magnetic field, giving rise to `colossal' values of the magnetoresistance (CMR) for some compositions \cite{Dagotto2001}. 

\begin{figure}
\includegraphics[width=8.5cm]{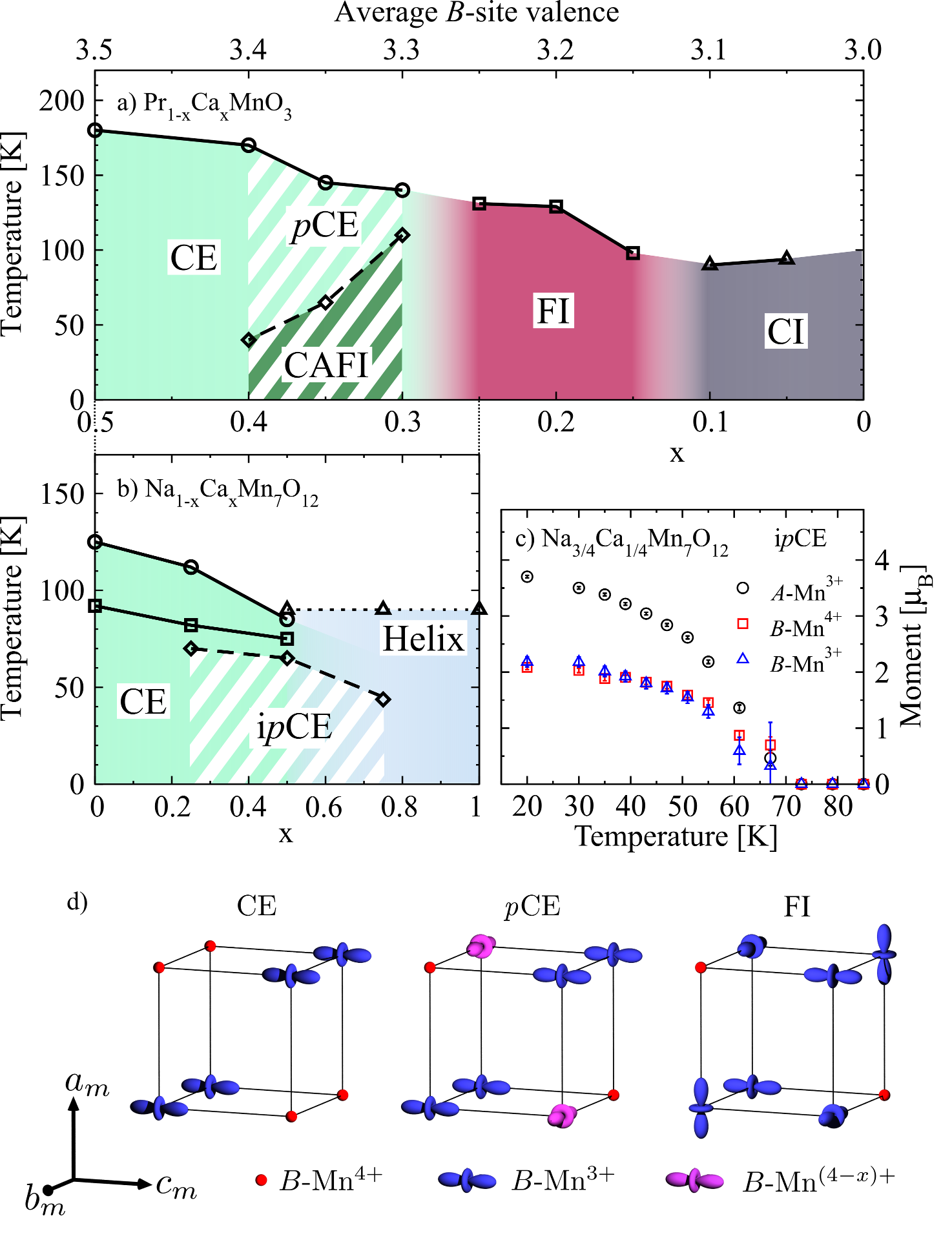}
\caption{\label{fig:all_phases}a) Pr$_{1-x}$Ca$_x$MnO$_3$ phase diagram adapted from reference \citenum{tomioka96}. b) \ncmox\ phase diagram based upon the present study. Both (a) and (b) have a common average $B$ site valence scale. c) Temperature dependence of magnetic moments that compose the i$p$CE structure of \ncmo. d) Cartoons of $B$ site orbital ordering that spans average $B$ site valences from +3.5 to +3.25. Note that the purple $B$-Mn$^{(4-x)+}$ orbitals indicate a likely orbital filling scheme.}
\end{figure}

More recently, the family of \emph{quadruple} perovskites has attracted significant interest \cite{Zeng99,prodi04,Vasilev07,Johnson12,Gilioli14,Belik16_1}. In these systems the $A$ sites have a 3/4 ordered occupation of transition metal ions, which is accommodated by large distortions of the perovskite structure. When the $A$ site transition metal is Mn$^{3+}$ the chemical formula can be written $A^{1+}_{1-x}A^{2+}_x$Mn$^{3+}_3$Mn$_4$O$_{12}$, such that the average valence of Mn on the $B$ site is $3.5-x/4$, which spans an analogous portion of the doping range of the simple perovskites (Fig. 1b). In the $x=0$ end member NaMn$_7$O$_{12}$, the average $B$ site valence is the same (+3.5) as for the famous half-doped simple perovskite La$_{0.5}$Ca$_{0.5}$MnO$_3$ \cite{Radaelli1997,prodi04}. As a result, the two compounds display very similar $B$ site charge, orbital and magnetic ordering - the well-known CE structure \cite{Goodenough1955}. In the quadruple perovskite the $A$ site Mn$^{3+}$ ions also form a magnetic sublattice, which independently adopts a different commensurate order at lower temperature \cite{prodi04}. By contrast, the $x=1$ end-member $A^{2+}$Mn$_7$O$_{12}$ with $A$=Ca, Pb, Cd..., isovalent with La$_{0.75}$Ca$_{0.25}$MnO$_3$, has thus far been regarded as unrelated to the magnetic and orbitally ordered phases observed in the simple perovskites, since it displays an \emph{incommensurate} magneto-orbital helix \cite{Johnson12, perks2012, johnson17}.

In this letter, we present the discovery of an incommensurate antiferromagnetic structure in \ncmox, which represents the `missing link' between simple and quadruple perovskites. In this new magnetic structure, labelled `i$p$CE', $B$ site spin ordering is locally identical to that of the commensurate and collinear \emph{pseudo}-CE ($p$CE) phase found in, for example, Pr$_{0.7}$Ca$_{0,3}$MnO$_3$ \cite{yoshizawa95}, but with a long range incommensurate spin rotation superimposed. We show that incommensurate order naturally arises in the quadruple perovskites through magnetic exchange between $A$ and $B$ sites above a critical value of $x$, and that the i$p$CE phase can transition into the CE structure of \nmo, or into the helical structure of \cmo, by changes in orbital order induced by electron doping - in full analogy with the simple perovskites (Figure \ref{fig:all_phases}). More generally, the $A$-$B$ exchange provides the handle that unifies the single and quadruple perovskite phase diagrams, generating distinct magnetic phases with common charge and orbital ordering.

Powder samples of \ncmox, with $x=0.25$, 0.5, and 0.75, were synthesised in high pressure/high temperature conditions using precursors obtained through the sol-gel route, as described elsewhere \cite{gilioli05}. The $x=0.5$, and 0.75 samples were encapsulated in a gold foil and hydrostatically pressed to 5 GPa, then heated at 800$^\circ$C, followed by quenching after 2 hours. The sodium-rich member of the series ($x=0.25$) required platinum foil encapsulation and increased pressure and temperature conditions; $P$=6.5 GPa, $T$=1100 $^\circ$C for 1.5 hours of reaction time. Neutron powder diffraction (NPD) measurements were performed on the WISH time-of-flight diffractometer \cite{Chapon11} at ISIS, the UK Neutron and Muon Spallation Source. The three samples were loaded into cylindrical vanadium cans and mounted within a $^4$He cryostat. Data were collected with high counting statistics in each magnetic phase, and with lower counting statistics on warming through all magnetic phases. The diffraction data were refined using \textsc{fullprof} \cite{Rodriguezcarvaja93}. In the following we focus on \ncmo. Details of the $x=0.5$ and 0.75 data analyses are given in the Supplemental Material \cite{SM}.

\begin{figure}
\includegraphics[width=8.5cm]{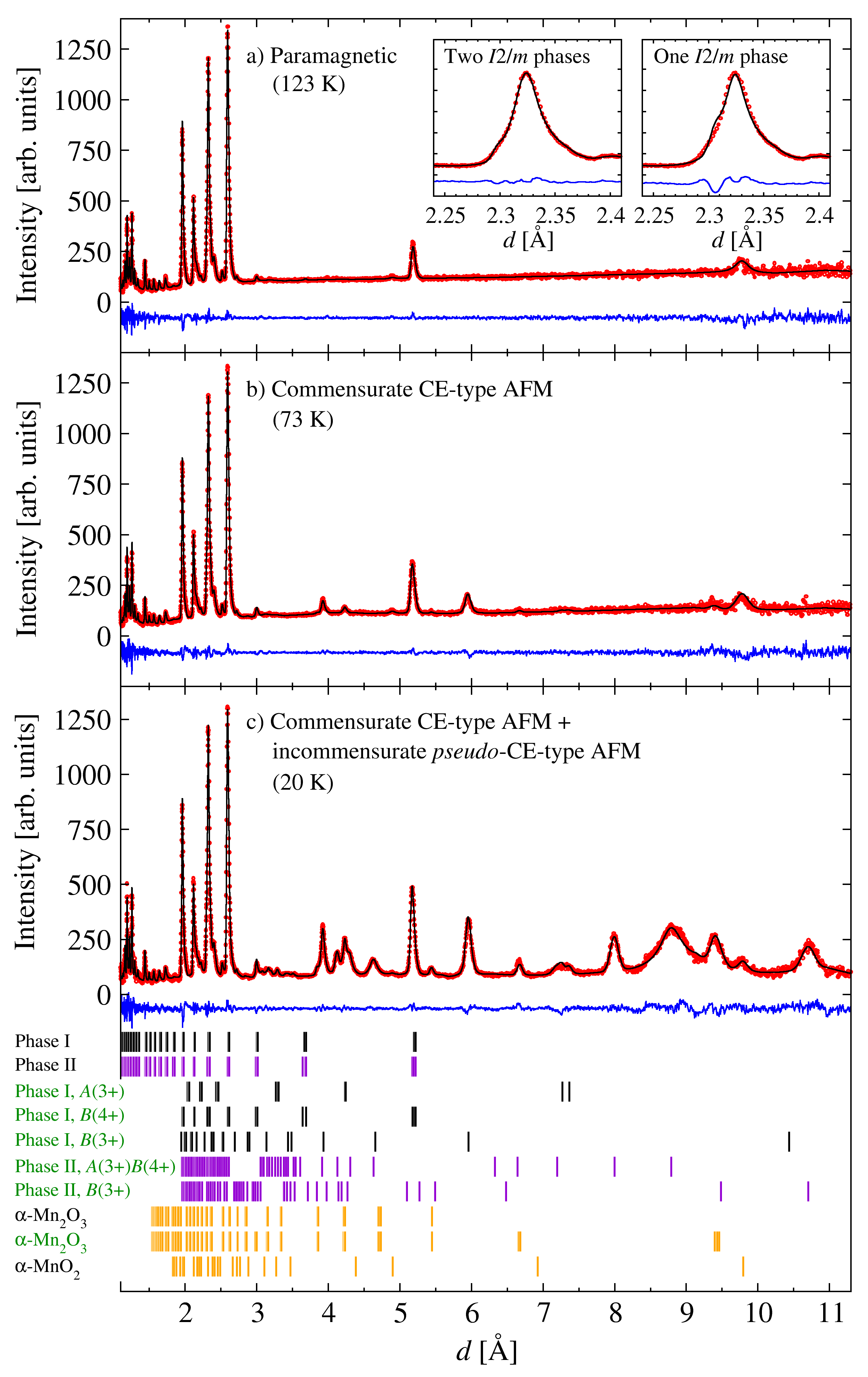}
\caption{\label{fig:npd}Neutron powder diffraction data (red points) measured in bank 2 (average $2\theta = 58.3^\circ$) of the WISH diffractometer from a) the paramagnetic, b) CE, and c) CE + $p$CE magnetic phases of \ncmo. The fitted models described in the main text are shown as solid black lines. Difference patterns ($I_\mathrm{obs} - I_\mathrm{calc}$) are given as solid blue lines. The location of nuclear (black label) and magnetic (green label) diffraction peaks from Phase I (black ticks), Phase II (purple ticks), and impurity phases (orange ticks) are given at the bottom of the figure. The inset to (a) highlights the presence of two $I2/m$ nuclear phases, I and II.}
\end{figure}

At room temperature the \ncmo\ diffraction pattern comprised a single, cubic ($Im\bar{3}$) quadruple perovskite phase, as observed in \nmo\ \cite{prodi04}, and $\sim5$ wt$\%$ $\alpha$-Mn$_2$O$_3$ and $\alpha$-MnO$_2$ impurity phases. Upon cooling, a structural phase transition was observed at $\sim160$ K, broadly consistent with the onset of charge and orbital order observed in \nmo\ below 175 K \cite{prodi04}. At 123 K (Figure \ref{fig:npd}a), the lineshape of the brightest diffraction peaks could only be fit by a model that included two, coexisting monoclinic phases in the ratio 44$\%$ : 56$\%$. Such phase coexistence is common in the simple perovskites \cite{tokura06}, and is consistent with discrete phase boundaries between different orbitally ordered states. Our data were fitted using the low temperature charge-ordered $I2/m$ crystal structure of \nmo\ \cite{prodi04}, disregarding weak superstructure reflections due to orbital ordering that index with propagation vector $(0.5,0,-0.5)$ \cite{streltsov14, prodi14}. The lattice parameters of the coexisting phases at 123 K were refined to $a=7.353(2)$, $b=7.2583(7)$, $c=7.340(2)$, $\beta = 90.31(1)$, and $a=7.350(1)$, $b=7.304(1)$, $c=7.332(2)$, $\beta = 90.10(3)^\circ$. The most significant difference between the two sets of parameters is the length of the $b$-axis. In the following, we label the structure with shortest $b$-axis `Phase I', and that with longest $b$-axis `Phase II'. 

Magnetic diffraction peaks appeared on cooling below 115 K, which could be indexed with propagation vectors $(0,0,0)$ and $(0.5,0,-0.5)$ with respect to the reciprocal lattice of Phase I. A second set of magnetic peaks appeared below 80 K, indexing with propagation vector $(0,1,0)$, also with respect to Phase I.  Refinements against data measured at 73 K (Figure \ref{fig:npd}b, $R_\mathrm{mag} = 6.5\%$) unambiguously showed that these peaks originate from a collinear CE magnetic structure, as observed in \nmo\ \cite{prodi04}, where the $B$ site sublattice orders at a higher temperature than the $A$ site sublattice. Appearing simultaneously below 70 K, a third set of magnetic diffraction peaks could be indexed with respect to Phase II using two incommensurate propagation vectors, $\mathbf{k}_1 = (k_x,1,k_z)$ and $\mathbf{k}_2 = (k_x+0.5,1,k_z-0.5)$, where $k_x=0.1606(3)$, and  $k_z=-0.0958(2)$ (these values were found to be temperature independent). In the commensurate limit  ($k_x=k_z=0$), these propagation vectors could describe a CE-type magnetic structure, but with \emph{ferromagnetic} interlayer coupling - corresponding to the so-called $p$CE structure in the simple perovskites \cite{yoshizawa95}.  In this case, $\mathbf{k}_1$ would be assigned to the \amthree\ and \bmfour\ sublattices, while $\mathbf{k}_2$ to the \bmthree\ sublattice, which was therefore used as a starting point for subsequent refinements.

\begin{figure}
\includegraphics[width=8.0cm]{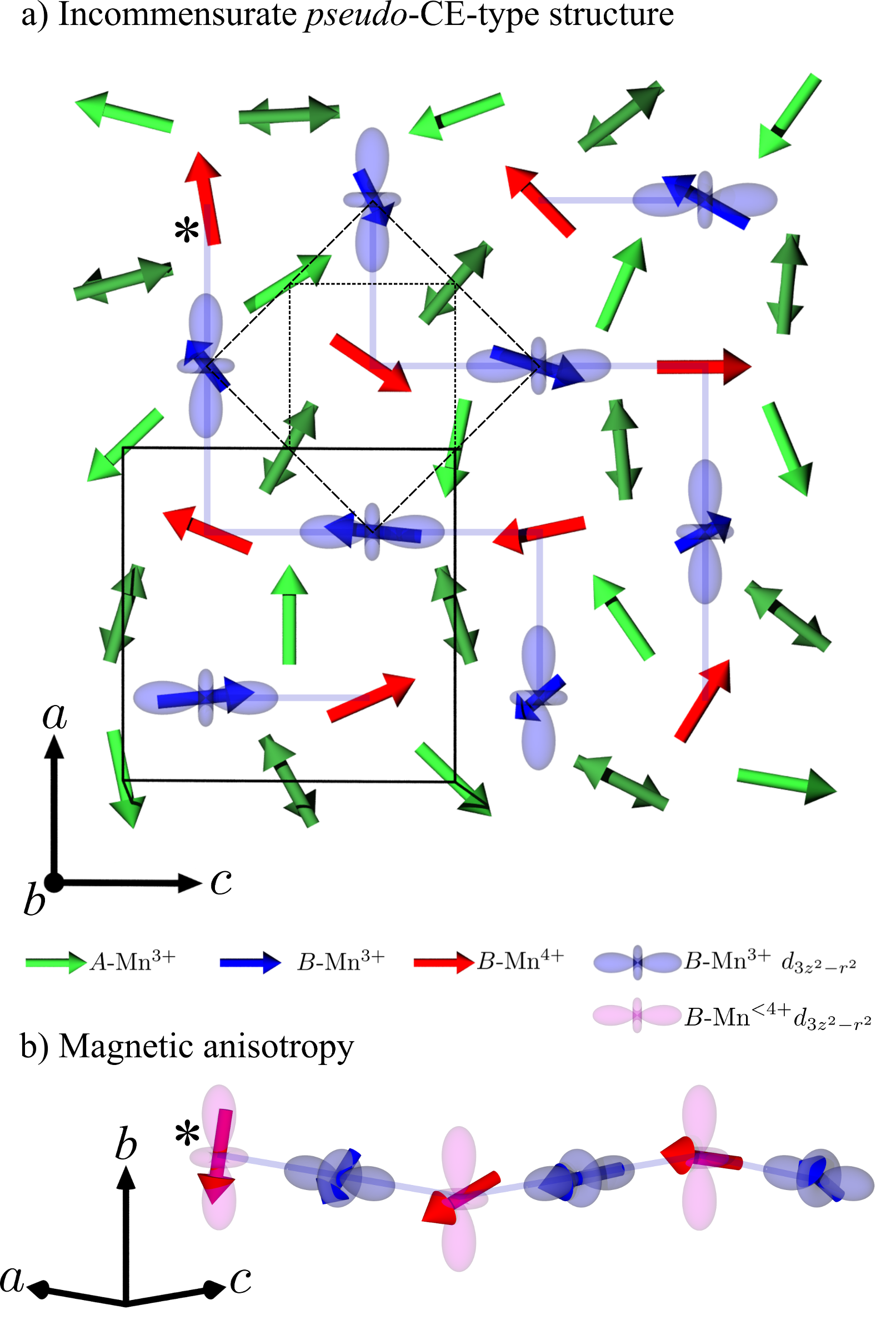}
\caption{\label{fig:mag_structure}a) The i$p$CE magentic structure with magnetic moments artificially shown to rotate in the $ac$-plane. The origin of the incommensurate modulation has been placed at the centre of the zeroth unit cell, drawn as a black solid line. For reference, the simple perovskite primitive unit cell and orbital supercell are superimposed, drawn as dotted and dashed lines, respectively. Double headed arrows denote $b$-axis stacking of moments that are antialigned within the same unit cell. Single headed arrows denote ferromagnetic stacking along $b$. b) The orbital structure of the zig-zag chains, with spins rotating in the true $bc$ plane. \amthree, \bmfour, and \bmthree\ moments are colored green, red, and blue, respectively. \bmthree\ and \bmfour\ $d_\mathrm{3z^2-r^2}$ orbitals are shaded in blue and pink, respectively.}
\end{figure}

An incommensurate $p$CE-type magnetic structure model (labeled i$p$CE) with magnetic moments rotating in the $bc$ plane, as described below, was fit to NPD data measured at 20 K. The model was found to be in excellent agreement with the data (Figure \ref{fig:npd}c, $R_\mathrm{mag} = 8.0\%$). Other moment rotation planes were tested, but they did not faithfully reproduce the data. Magnetic diffraction peaks that, by geometry, were primarily sensitive to the component of long range magnetic order orthogonal to the $ac$ plane were found to be anomalously broad, which might indicate a short coherence length along the $b$ axis due to antiferromagnetic stacking faults. The i$p$CE magnetic structure is illustrated in Figure \ref{fig:mag_structure}a, where spins have been artificially rotated into the $ac$-plane to provide a perceptual overview of the rotating structure. Figure \ref{fig:mag_structure}b shows a perspective view of the magnetic structure along a zig-zag chain (blue line in pane a) with spins in their true orientation.

Except for the incommensurate spin rotation, the \amthree\ sublattice magnetic order is the same as in \nmo, with every nearest neighbour moment aligned antiferromagnetically. The $ac$-plane magnetic structure of the $B$ sites is also similar to the CE phase of \nmo, with ferromagnetic zig-zag chains running along the $[\bar{1}01]$ direction (blue lines in Figure \ref{fig:mag_structure}) and coupled antiferromagnetically with each other. However, the coupling in the $b$ direction is \emph{ferromagnetic}, as for the $p$CE structure. This difference can be interpreted exactly as in the case of simple perovskites: at half doping, all the $d_{3y^2-r^2}$ orbitals providing coupling in the $b$ direction are unoccupied, yielding interlayer antiferromagnetic superexchange of the CE phase \cite{asaka02, prodi04}. As the nominal Mn$^{4+}$ site becomes progressively occupied by Mn$^{3+}$ with its half-filled $d_{3y^2-r^2}$ orbital oriented approximately along $b$, ferromagnetic exchange increasingly competes with and eventually supplants antiferromagnetism at a critical value of doping, giving rise to the $p$CE-type phase \cite{jirak85}. We note that the observed expansion of the $b$ lattice parameter of Phase II with respect to Phase I is consistent with such additional orbital occupation upon increased electron doping. It is also noteworthy that the occupation of $d_{3y^2-r^2}$ orbitals along $b$ reduces the on-site easy-$ac$-plane spin anisotropy, which would otherwise disfavour the observed $ab$ spin rotation plane (Figure \ref{fig:mag_structure}b).

NPD data measured from Na$_{0.5}$Ca$_{0.5}$Mn$_7$O$_{12}$ and Na$_{0.25}$Ca$_{0.75}$Mn$_7$O$_{12}$ were fit using the same method described above. In both cases the i$p$CE phase was present and ordered with the same values of $k_x$ and $k_z$. Moreover, an additional helical phase had to be included (see Supplemental Material - Sections S1 to S4 \cite{SM}). From these analyses the phase diagram for \ncmox\ was constructed (Figure \ref{fig:all_phases}b) and matched to the average $B$-site valence of Pr$_{1-x}$Ca$_x$MnO$_3$ (Figure \ref{fig:all_phases}a). One finds a close correspondence between the two phase diagrams, providing a unified picture of charge, orbital, and magnetic ordering in the simple and quadruple perovskite manganites.  This similarity is further strengthened by the observation that, in the commensurate limit, the orbital component of the $A^{2+}$Mn$_7$O$_{12}$ helical phase \cite{johnson17} is analogous to the orbital polaron lattice proposed for the ferromagnetic insulating (FI) phase of Pr$_{1-x}$Ca$_x$MnO$_3$ \cite{mizokawa00}.

We now consider why the $p$CE-type magnetic structure is incommensurate in the quadruple perovskites, yet commensurate in the simple perovskites, while the CE-type structures are commensurate in both. In all phases, the \bmthree\ magnetic moments are coupled to the \bmfour\ moments via the orbital order.  In fact, at the phenomenological level a trilinear free energy invariant can be constructed that couples the respective order parameters (Supplemental Material, Section S5 \cite{SM}). However, a key difference between the simple and quadruple perovskites is the $A$-site magnetism. Firstly, one can trivially observe that in the CE and $p$CE phases of the simple perovskites the $A$ and $B$ sites are magnetically decoupled as the $A$ site sublattice is non-magnetic. In the CE structure of the quadruple perovskite the $A$ and $B$ site magnetic orderings are also decoupled, because the respective propagation vectors are different and no phenomenological coupling term that is quadratic in the spins can therefore exist. Confirmation of this is found in the fact that the $A$ site and $B$ site phase transition temperatures are different. By contrast, in the $p$CE phase the \amthree\ and \bmfour sublattices order with the same propagation vector, while the latter is coupled to \bmthree through the orbital ordering as before. Hence, all three sites are in principle allowed to couple. Indeed, the temperature dependence of the magnetic moment magnitudes of \amthree, \bmfour, and \bmthree\ for the i$p$CE phase of \ncmo, shown in Figure \ref{fig:all_phases}c, demonstrates a \emph{single} phase transition at 70 K due to coupling between all magnetic sublattices. 

Crucially, by calculating the mean-field Heisenberg exchange energy of a generic CE/$p$CE-type magnetic structure, parametrised by the magnetic propagation vector, one can show that the $A$-$B$ coupling term is exactly zero in the commensurate case. This approach also enables a magnetic phase diagram to be constructed as a function of the effective exchange interactions. The calculation of our exchange model is developed in the Supplemental Material - Section S6 \cite{SM}.  Here, we only present the final result for the energy difference between $p$CE-type and CE phases:

\begin{eqnarray}
\label{eq: energy_difference}
E_{pCE}-E_{CE}&=&\tilde{J}( \cos{\alpha} \cos{\beta}-1)-\Delta \tilde{J} \sin \alpha \sin \beta\nonumber\\
&&   -2 J_y^{BB} - J^{AB} \sin \alpha 
\end{eqnarray}
where $\alpha=\tfrac{\pi}{2}(k_x-k_z)$ and $\beta=\tfrac{\pi}{2}(k_x+k_z)$.  $\tilde{J}$ and $\Delta \tilde{J}$ (defined in the Supplemental Material) are linear combinations of exchange parameters for $A$-$A$ interactions and for $B$-$B$ interactions in the $xz$ plane, with $\Delta \tilde{J}$ defining the in-plane anisotropy. $J_y^{BB}$ is the average exchange between $B$ sites separated along the $b$ axis, while $J^{AB}$ is a net $A$-$B$ exchange parameter.  One can immediately see from Equation \ref{eq: energy_difference} that the $A$-$B$ interaction term is \emph{zero} for a commensurate structure.  In fact, one can show that the corresponding coupling term is zero by symmetry (see Supplemental Material), and that this result does not therefore depend on any approximation we have introduced in our model. 

\begin{figure}
\includegraphics[width=8.5cm]{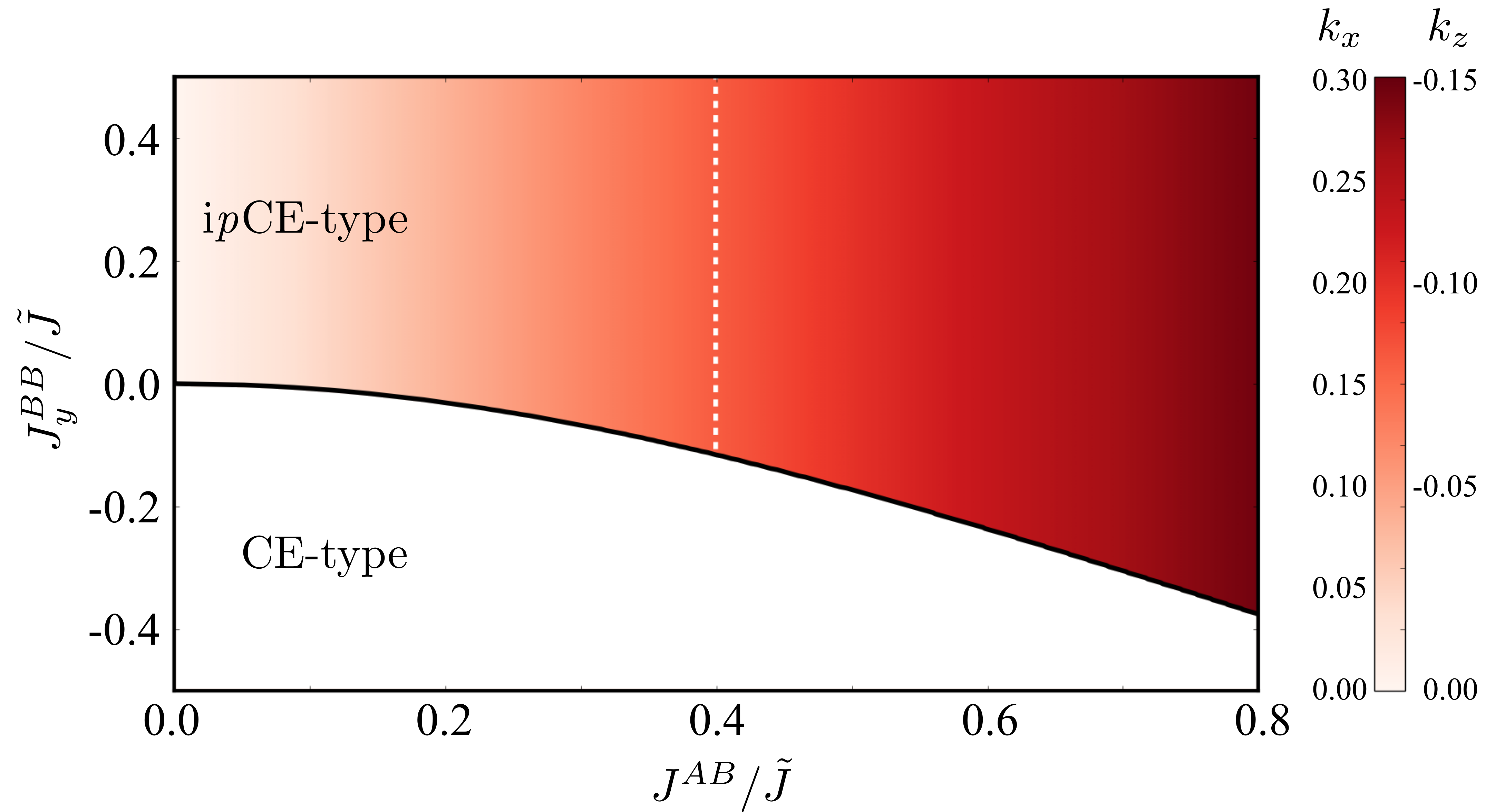}
\caption{\label{fig:ncmo_phases}Magnetic phase diagram as determined from model Heisenberg exchange energy calculations described in the main text, and presented in detail in the Supplemental Material \cite{SM}. The vertical dashed line demarks the value of $J^{AB}/\tilde{J}$ determined from the experimental values of $k_x$ and $k_z$ found in all measured i$p$CE phases.}
\end{figure}

Eq. \ref{eq: energy_difference} can be employed to construct the ground-state phase diagram, which is 3-dimensional in the parameters $\Delta \tilde{J}/ \tilde{J}$,  $J_y^{BB}/ \tilde{J}$ and $J^{AB}/ \tilde{J}$:  the minima of eq. \ref{eq: energy_difference} can be found analytically (see Supplementary Information), and corresponds to the stable mean-field values of the propagation vector \emph{if} $E_{pCE}-E_{CE}<0$, otherwise the commensurate CE phase is stabilized. Figure \ref{fig:ncmo_phases} depicts a 2-dimensional section of the 3-dimensional phase diagram, in which the parameter $\Delta \tilde{J}/ \tilde{J}$ has been fixed to the value determined from the experimental propagation vectors.  For a non-magnetic $A$ site or when $J^{AB}=0$, the the CE and $p$CE phases are both commensurate and the transition between them occurs when $J_y^{BB}$ changes sign from antiferromagnetic to ferromagnetic.  By contrast, a non-zero $J^{AB}$ favours the i$p$CE phase, which therefore becomes stable even for weakly antiferromagnetic $J_y^{BB}$.  Moreover, the i$p$CE phase is always incommensurate for non-zero $J^{AB}$.

We note that our mean field model is derived only for the charge and orbital order pertinent to the CE, $p$CE, and i$p$CE phases, and does not include, for example, the helical magnetic phase of CaMn$_7$O$_{12}$, nor the FI phase of Pr$_{1-x}$Ca$_x$MnO$_3$, which both share charge and orbital orders different to the former phases. However, the same methodology could be used to construct a more general
 mean field model, in which the quadruple perovskites support magnetic structures that are incommensurate as a result of frustration introduced by the presence of an ordered pattern of magnetic and non-magnetic $A$ sites \cite{perks2012,johnson17}. A complete treatment of a generalised Heisenberg exchange model will be the subject of a future work.

In summary, we have discovered a new antiferromagnetic structure in the quadruple perovskite \ncmo\ - an incommensurate \emph{pseudo}CE-type structure - which can be thought of as the `missing link' connecting the charge, orbital and magnetic ordering in Mn$^{3+}$-rich `canonical' $A$MnO$_3$ simple perovskites, and the $A$Mn$_7$O$_{12}$ quadruple perovskites. The observed magnetic structures can be understood as mean field solutions of a Heisenberg Hamiltonian tuned by orbital ordering and symmetry. We have shown that noncollinear incommensurate magnetic order is stabilized in the quadruple perovskite by magnetic interactions between $A$ and $B$ sites. This fact also explains why the magnetic structures of simple perovskites are generally collinear and commensurate, as in this case the $A$ site sublattice is non-magnetic. Together, these findings motivate the development of a truly generalised mean-field model of magneto-orbital ordering that captures not only the FI and helical phases, but also other phases such as rare earth quadruple perovskite analogues of LaMnO$_3$. Experimentally, the combined phase diagrams can be expanded to other thermodynamic variables, such as high magnetic fields and hydrostatic pressure, in addition to a broader search for other quadruple perovskites with highly significant properties, such as metal-to-insulator transitions and colossal magneto-resistance.

\begin{acknowledgments}
RDJ acknowledges support from a Royal Society University Research Fellowship. PGR and RDJ acknowledge support from EPSRC, grant number EP/M020517/1, entitled ``Oxford Quantum Materials Platform Grant''.
\end{acknowledgments}

\bibliography{ncmo}

\end{document}